\documentstyle[12pt]{article}

\catcode`\*=11  
\def\slashsym#1#2{\mathpalette{\sl*sh{#1}}{#2}}
\def\sl*sh#1#2#3{\ooalign{\setbox0=\hbox{$#2\not$}
                          $\hfil#2\mkern-24mu\mkern#1mu
                           \raise.15\ht0\box0\hfil$\cr
                          $#2#3$}}
\catcode`\*=12
\newcommand{\beq}{\begin{equation}}
\newcommand{\eeq}{\end{equation}}
\newcommand{\beqa}{\begin{eqnarray}}
\newcommand{\eeqa}{\end{eqnarray}}

\newcommand{\eq}[1]{(\ref{#1})}
\newcommand{\ket}[1]{\vert\,{#1}\,\rangle}
\newcommand{\bra}[1]{\langle\,{#1}\,\vert}
\newcommand{\ra}{\rightarrow}
\newcommand{\del}{\partial}
\newcommand{\tr}{{\rm Tr}}
\newcommand{\sh}{{\rm sh}}

\newcommand{\teta}{\Theta_2(s{2\mu\over\beta},is{4\pi\over\beta^2})}
\newcommand{\integ}{\int_0^\infty}
\newcommand{\rinv}{{2\over R}}
\newcommand{\xy}{\langle\,x(s)\,\ket{y(0)} }
\newcommand{\NP}[1]{ {\it Nucl.~Phys.} {\bf #1}}
\newcommand{\PL}[1]{ {\it Phys.~Lett.} {\bf #1}}
\newcommand{\Prep}[1]{ {\it Phys.~Rep.} {\bf #1}}
\newcommand{\PR}[1]{ {\it Phys.~Rev.} {\bf #1}}

\newcommand{\PTP}[1]{ {\it Prog.~Theor.~Phys.} {\bf #1}}

\newcommand{\MPL}[1]{ {\it Mod.~Phys.~Lett.} {\bf #1}}
\newcommand{\IJMP}[1]{ {\it Int.~Jour.~Mod.~Phys.} {\bf #1}}

\newcommand{\ZP}[1]{ {\it Z.~Phys.} {\bf #1}}
\begin{document}
\topmargin 0pt
\oddsidemargin 1mm
\begin{titlepage}

\begin{flushright}
OU-HET 226  \\
NBI-HE-95-38\\
hep-th/9511059\\
\end{flushright}
\setcounter{page}{0}
\vspace{11mm}
\begin{center}
{\Large \bf{ Approach to $D$-dimensional Gross-Neveu Model \\
          at Finite Temperature and Curvature\\} }
\vspace{20mm}

{\large \bf{Shinya Kanemura}
\footnote{ kanemu@phys.wani.osaka-u.ac-jp}}\\
{\em Department of Physics, Osaka University\\
     Machikaneyama 1-16, Toyonaka, Osaka 560, Japan}\\
and\\
{\large \bf{Haru-Tada Sato}
\footnote{ Fellow of the Danish Research Academy,\,\,\,
 sato@nbivax.nbi.dk}}\\
{\em The Niels Bohr Institute, University of Copenhagen\\
     Blegdamsvej 17, DK-2100 Copenhagen, Denmark}\\
\end{center}
\vspace{7mm}

\begin{abstract}
We discuss phase structure of chiral symmetry breaking of the $D$-dimensional
($2\leq D\leq3$) Gross-Neveu model at finite temperature, density and constant
curvature. We evaluate the effective potential in a weak background
approximation to thermalize the model as well as in the leading order of the
$1/N$-expansion. A third order critical line is observed similarly to the
$D=2$ case.
\end{abstract}

\vspace{1cm}

\end{titlepage}
\newpage
\renewcommand{\thefootnote}{\arabic{footnote}}
\indent

Finite temperature field theories with a curvature background may be useful
for investigation of a simultaneous effect from high temperature and strong
background under certain circumstance like an early universe \cite{book}.
However, there seems few convenient way to handle both temperature
and background simultaneously other than lattice simulations and high/low
temperature expansions. In previous paper \cite{KS}, we showed a method to
introduce temperature under constant curvature background in two dimensions
considering large $N$ leading terms of the Gross-Neveu model \cite{GN}.
Although our thermalization procedure is valid for weak curvature cases, it
still works well even in strong curvature regions if temperature is
small. The chiral phase transitions of the Gross-Neveu model in terms of
temperature \cite{Tc} and of background curvature \cite{BK},\cite{Rc},
\cite{muta2} take place in the regions that temperature is sufficiently
larger than the square root of curvature for the former and the reverse is
the latter. The advantage of our method is to evaluate the simultaneous effect
to some extent in the $R$-$T$-$\mu$ phase space and thus to interpolate these
two regions each other.

In this paper, we apply the method of the previous analysis to the effective
potential of the $D$-dimensional discrete chiral Gross-Neveu model defined in
the leading term of large $N$ expansion and show its phase diagrams in
$D=2.5$ and $D=3$. One of our aims is to confirm that our method is valid
not only for a weak curvature side $R\ll T$ but also for a low temperature
side $T\ll R$ in arbitrary dimension $2\leq D<4$.

Let us derive the effective potential in a suitable form to apply our method.
The Notation of this paper follows \cite{BK} and \cite{KS}.
The effective potential under a constant curvature is obtained in the same
way as \cite{BK} using the proper time method:
\beq
V(\sigma;R)={1\over2\lambda}\sigma^2
           -i{\rm tr}[1]\int_0^\sigma dm S(x,x;m),        \label{eq1}
\eeq
where ${\rm tr}[1]$ means the dimension of gamma matrices and
\beq
S(x,y;m)=-i\integ ds e^{-ism^2}
         \bra{x(s)} {\slashsym6\pi}+m \ket{y(0)}          \label{eq2}
\eeq
is evaluated through the proper time Hamiltonian
$H=-\pi^2+{R\over8}\tr(\sigma^2)$ in the following way.
The Schr{\"o}dinger equation for the quantity $\xy$ is
\beq
-i{\del\over\del s} \xy=
\left[(x-y)K(x-y)+i{R\over8}\tr(\sigma \cot{R\over4}\sigma s)
          +{R\over8}\tr\sigma^2 \right] \xy               \label{eq3}
\eeq
with
\beq
K_{\mu\nu}={R^2\over64}
      \left(\sigma\over \sin({R\over4}\sigma s)\right)^2_{\mu\nu} \label{eq4}.
\eeq
The solution is given by
\[
\xy = (4\pi s)^{-D/2}\exp\left[-\int_y^x d\xi\Gamma(\xi)\right]
      \exp\left[-{i\over4}(x-y){R\sigma\over4}
       \cot({R\sigma s\over4})(x-y)\right]
\]\beq
\times \exp\left[ i{Rs\over8}\tr(\sigma^2) \right]
       \exp \left[-{1\over2}\tr\ln
          \left({\sin(Rs\sigma/4)\over Rs\sigma/4}\right)\right],  \label{eq5}
\eeq
and
\beq
\bra{x(s)} \pi_{\mu} \ket{y(0)}= i{R\over8}\sigma
 \left\{ 1-i\cot({R\sigma s\over4}) \right\}(x-y)\xy,
\label{eq6}
\eeq
where $\Gamma(\xi)$ is a product of spin connections and $\sigma^{\mu\nu}$
matrices and
\beq
  \exp \left\{ -{1\over2}\tr\ln
              \left({\sin(Rs\sigma/4)\over Rs\sigma/4}\right) \right\}
= \left( {Rs(D-1)\over4\sin(Rs(D-1)/4)} \right)^{1/2}
  \left( {Rs\over4\sin(Rs/4)} \right)^{D-1\over2}.               \label{eq7}
\eeq
Here we have used a naive Taylor expansion for the logarithmic function
and the following formula
\beq
\tr(\sigma^n) = (D-1)^n + (-1)^n(D-1).                \label{eq8}
\eeq
It can be easily checked that the equation \eq{eq3} is satisfied assuming
the relation
\beq
\tr\left[\sigma \cot({Rs\sigma\over4})\right]
             =\cot{Rs\over4}+\cot{Rs(D-1)\over4}.\label{eq9}
\eeq
Rotating $s \ra is$, we obtain
\beq
S(x,x;m)=-im\integ {ds\over(4\pi s)^{D/2}}e^{-s \left\{m^2+{R\over8}D(D-1)
\right\} }
\left( {Rs(D-1)/4\over \sh(Rs(D-1)/4)} \right)^{1/2}
  \left( {Rs/4\over \sh(Rs/4)} \right)^{D-1\over2},       \label{eq10}
\eeq
which precisely reproduces the previous $D=2$ case. For later convenience,
we use a rescaled curvature as $R \ra 2R/D(D-1)$ hereafter and define
\beq
F(s;D)\equiv
             \left( { s/D\over \sh(s/D)} \right)^{1/2}
  \left( {s/D(D-1)\over \sh(s/D(D-1))} \right)^{D-1\over2}.
\eeq
Using the formula
\beq
(4\pi s)^{-D/2} = \int_{-\infty}^{\infty}
                       {d^Dk\over(2\pi)^D}e^{-sk^2},         \label{eq11}
\eeq
the bare potential reads
\beq
V(\sigma;R)={1\over2\lambda}\sigma^2+{1\over2}{\rm tr}[1]
            \int{d^Dk\over(2\pi)^D}\integ{ds\over s}F({R\over2}s;D)
            e^{-s(k^2+R/4)}(e^{-s\sigma^2}-1).        \label{eq12}
\eeq
With the replacement which corresponds to a weak field approximation,
\beq
k^2\quad\ra\quad({2n+1\over\beta}\pi-i\mu)^2+k_i^2,\hskip 30pt
\int{dk_0\over2\pi}\quad\ra\quad{1\over\beta}\sum_n,        \label{eq13}
\eeq
and with integration over $k_i$ and summation over $n$, the bare potential
with finite $R$,$\beta$,$\mu$ parameters of our approach becomes
\beq
V(\sigma;R,\beta,\mu)={1\over2\lambda}\sigma^2+{{\rm tr}[1]\over2\beta}
            \integ {ds\over s}{\teta\over(4\pi s)^{(D-1)/2}}F({R\over2}s;D)
            e^{-s(R/4-\mu^2)} (e^{-s\sigma^2}-1),    \label{eq14}
\eeq
where $\Theta_2$ is the elliptic theta function. If $R$ is small, the thermal
function $\Theta_2$ is dominant. Conversely for small $T$, the curvature
function $F$ is dominant. Extending $R$ and $T$ to finite regions,
we may catch a glimpse of phase structure. Now taking the same renormalization
procedure as previous paper, the renormalized coupling constant $\lambda_R$ is
thereby
\beq
{1\over\lambda}-{1\over\lambda_R}={\rm tr}[1]\integ {ds\over
       (4\pi s)^{D/2}}e^{-s(1+R/4)}(1-2s)F({R\over2}s;D),     \label{eq15}
\eeq
and the renormalized effective potential of our model is therefore
\[
\hskip -55pt
V(\sigma;R,\beta,\mu)={1\over2\lambda_R}\sigma^2+{1\over2}{\rm tr}[1]\integ
{ds\over(4\pi s)^{D/2}} e^{-sR/4} F({R\over2}s;D)
\]
\beq
\hskip 35pt
\times \left\{,{1\over s}(e^{-s\sigma^2}-1){\sqrt{4\pi s}\over\beta}
e^{s\mu^2}\teta+\sigma^2e^{-s}(1-2s)\,\right\}.                \label{eq16}
\eeq

The gap equation for a dynamical mass and the critical surface for second
order phase transitions are respectively
\beq
0={1\over\lambda_R}+{\rm tr}[1]\integ {ds\over(4\pi s)^{D/2}}
e^{-sR/4} F({R\over2}s;D)
\left\{\,-e^{-s(\sigma^2-\mu^2)}{\sqrt{4\pi s}\over\beta}\teta +
                        e^{-s}(1-2s)\,\right\},                 \label{eq17}
\eeq
and
\beq
0=\integ{ds\over(4\pi s)^{D/2}} \left[\, e^{-sR/4} F({R\over2}s;D)
  \left\{ -e^{s\mu^2}{\sqrt{4\pi s}\over\beta}\teta
  +  e^{-s}(1-2s) \right\}  + 2se^{-s} \,\right],                \label{eq18}
\eeq
where we have adopted the value of the renormalized coupling constant
\beq
{1\over\lambda_R}=
         {\rm tr}[1]\integ {ds\over(4\pi s)^{D/2}}\,2s\,e^{-s}, \label{eq19}
\eeq
which means the broken phase with a dynamical mass $\sigma=1$ at $T=\mu=R=0$.

Let us make some remarks on the equation \eq{eq18} of the second order
critical surface. First, note that \eq{eq18} is not always sitting on true
second order transitions if first order transitions occur. Similarly to
\cite{KS}, we need the following equation to determine tri-critical points
which divide the critical surface into the first and second order ones;
\beq
\lim_{\sigma\ra0}({\del\over\del\sigma^2})^2
                 V(\sigma;R,\beta,\mu) =0,              \label{eq206}
\eeq
namely
\beq
    0= \integ{ds\over(4\pi s)^{D/2}} se^{s(\mu^2-R/4)}F({R\over2}s;D)
         {\sqrt{4\pi s}\over\beta}\teta.      \label{eq207}
\eeq
The intersection between this surface \eq{eq207} and the second order
surface \eq{eq18} means just a tri-critical line.

Second, we can reproduce several known results
in particular limits. Considering the limit $R\ra0$ in \eq{eq18},
\beq
0=\integ{ds\over (4\pi s)^{D/2}}
\left\{\,
  e^{-s} -e^{s\mu^2}{\sqrt{4\pi s}\over\beta}\teta \,
\right\},  \label{eq20}
\eeq
and integrating, we obtain
\beq
\beta^{D-2}\Gamma(1-{D\over2})={2\over\sqrt{\pi}} (2\pi)^{D-2}
\Gamma({3-D\over2})\Re\zeta(3-D,{1\over2}+i{\beta\mu\over2\pi}),\label{eq21}
\eeq
where $\zeta$ is the generalized zeta function. This is exactly the equation
for the second order critical line on the $T$-$\mu$ plane derived in the
dimensional reguralization formalism \cite{Muta} as well as in the proper time
formalism for two-dimensional analysis \cite{KS}. It is shown in \cite{Muta}
that a variety of critical equations is reproduced \cite{creq},\cite{CM}.

Third, the limit $\beta\ra\infty$ in Eq.\eq{eq18}
\beq
0= \integ{ds\over(4\pi s)^{D/2}} \left[\, e^{-sR/4}F({R\over2}s;D)
  \{ -1 +  e^{-s}(1-2s) \}  + 2se^{-s} \,\right]               \label{eq22}
\eeq
becomes
\beq
    2\Gamma({D\over2})(\rinv)^{{D\over2}-1} +
            \psi_D(\rinv)+{4\over R}\psi_D'(1+\rinv)=0,    \label{eq23}
\eeq
where $\psi_D$ is
\beq
     \psi_D(z)=\integ ds s^{-D/2} e^{-s/2}(e^{-sz}-1)
              F(s;D),               \label{eq24}
\eeq
which is a generalization of the digamma function and
$\psi_D(z)=-\gamma-\psi(1+z)$ in two dimensions. Our formula precisely
reproduces Buchbinder-Kirillova's result as was mentioned in \cite{KS}.
One should note that our renormalization condition is slightly different
from that considered in \cite{muta2}.

Now let us discuss phase structure of the effective potential \eq{eq16}.
The method to find a critical surface is completely parallel to that of
the previous paper \cite{KS}. Let us make mention of critical lines on
the $R$-$\mu$ and $T$-$\mu$ planes.  The surface \eq{eq207} becomes the
following equation of line on the $R$-$\mu$ plane in $\beta\ra\infty$
\beq
    0=\integ {ds\over(4\pi s)^{D/2}}se^{-Rs/4}F({R\over2}s;D),  \label{eq28}
\eeq
however this line never intersect with the second order critical line
at least in the region $\mu\leq1$ on the $R$-$\mu$ plane. This means  no
tri-critical point in that region in arbitrary dimension $2\leq D <4$. On the
contrary, the limit $R\ra0$ of \eq{eq207}
\beq
  0=\integ {ds\over(4\pi s)^{D/2}}
          se^{s\mu^2}{\sqrt{4\pi s}\over\beta}\teta,     \label{eq29}
\eeq
coincides with the equation
\beq
       \Re \zeta(5-D,{1\over2}+i{\beta\mu\over2\pi})=0,     \label{eq30}
\eeq
and it is known that this gives a tri-critical point on $T$-$\mu$ plane
in $2 \leq D\leq 3$ \cite{Muta}.
These critical lines on each planes can be interpolated through chase to
the effective potential for first order transitions and to the critical
surface equation \eq{eq18} for second order transitions respectively.
We show in Fig.1 and Fig.2 the second order cases (represented by solid lines)
for $\mu=0$ and the first order cases (by dashed lines) for $\mu=0.8$ in each
dimensions $D=2,2.5$ and 3. In Fig.1, one can read not only that critical
points on the $T$ and $R$ axes are connected with each other but also that
the phase boundary of second order continuously spreads over as the dimension
$D$ increases. Similarly in Fig.2, we observe a continuous move of
tri-critical points as $D$ varies. Decreasing the value of $D$ from 3 or 4,
we first find one tri-critical point at certain value in $2.5<D<3$.
After then, we observe two tri-critical points at both edges
of a first order critical line (for example, see $D=2.5$ case in Fig.2).
Further decreasing $D$, the leftward tri-critical point lands on the $T$ axis
and then disappears. This move of tri-critical points may be clear
also from the tendency that the continuous move of phase
boundary as $D$ decreasing considerably resembles the case of $\mu$
increasing with fixed $D$. In other words, when we increase the value of $D$,
we thus encounter a similar critical line to that for smaller $\mu$ in
non-increased $D$.

In summary, we show critical lines in Fig.3 at various values of $\mu$ in
$D=3$. Differently from the cases $2\leq D<3$, there is no tri-critical point
on the $T$ axis excepting $T=0$ \cite{Kli}, however first order transitions
can be seen in the nonzero $R$ region similarly to $2\leq D<3$. We obtain two
tri-critical points ${\bf A}$ and ${\bf B}$
at $(R,T)=(0,0)$ and $(1.7,0.2)$ for $\mu=1$, respectively,
analyzing critical surface equations \eq{eq18} and \eq{eq207}. These
tri-critical points may be smoothly connected to each other along a first
order phase line shown in Fig.3. We actually find a first order transition
at $(1.5,0.1966)$ as an interpolation point. Unfortunately, we could not
trace this line any more because of a technical reason, {\it i.e.}, terribly
slow convergence of an integral for $R<1.5$ when $\mu=1$. This might be in a
range beyond approximation. However, the smooth connection is naturally
expected also from the tendency explained at the end of last paragraph.

For three dimensional models, existence of first order transitions by itself
is same feature as \cite{linear3} where the Riemann normal coordinate
expansion is employed and only linear curvature terms is evaluated (see also
4-dim. cases \cite{ELO} and \cite{linear4}), while our approach shows no first
order transition at some definite values of $\mu$ including $\mu=0$. In
particular for $\mu=0$, second order transition is in common with
\cite{muta2} in arbitrary dimensions $2\leq D<4$. Finally, we point out, as
far as concerned with our approximation, that certain finite values of $\mu$
are necessary to first order transitions triggered by a curvature background.

\vspace{1cm}
\noindent
{\em Acknowledgments}

The authors would like to thank T. Inagaki, K. Ishikawa, S. Mukaigawa, and
T. Muta for valuable suggestions and discussions.

\vspace{1cm}
\noindent
{\bf Note added}

We should note that Eqs.(23) and (28) of \cite{KS} are not valid for large
$\mu$. Instead, correct equations are obtained from this paper putting $D=2$.
However, one can immediately recognize that there is no essential differences
of the critical line on $R$-$\mu$ plane.
\newpage

%
\end{document}